**Spontaneous vortex state in a superconductor/ferromagnet nanocomposite**

Nanami Teramachi[1], Yusuke Seto[2], Takahiro Sakurai[3], Hithoshi Ohta[4] & Takashi Uchino[1*]

[1]Department of Chemistry, Graduate School of Science, Kobe University, Nada, Kobe 657-8501, Japan

[2]Department of Planetology, Graduate School of Science, Kobe University, Nada, Kobe 657-8501, Japan

[3]Center for Support to Research and Education Activities, Kobe University, Nada, Kobe 657-8501, Japan

[4]Molecular Photoscience Research Center, Kobe University, Nada, Kobe 657-8501, Japan

*e-mail: uchino@kobe-u.ac.jp

**The mechanism of the interplay between superconductivity and magnetism is one of the intriguing and challenging problems in physics[1–4]. Theory has predicted that the**

**ferromagnetic order can coexist with the superconducting order in the form of a spontaneous vortex phase in which magnetic vortices nucleate in the absence of an external field[1–7]. However, there has been no rigorous demonstration of spontaneous vortices by bulk magnetic measurements[4,8–10]. Here we show the results of experimental observations of spontaneous vortices using a superconductor/ferromagnet fractal nanocomposite, in which superconducting $MgB_2$ and ferromagnetic $\alpha'$-MnB nanograins[11] are dispersedly embedded in the normal Mg/MgO matrix to realize the remote electromagnetic interaction and also to induce a long-range Josephson coupling. We found from bulk magnetization measurements that the sample with nonzero remanent magnetization exhibits the magnetic behaviors which are fully consistent with a spontaneous vortex scenario predicted theoretically for magnetic inclusions in a superconducting material[7]. The resulting spontaneous vortex state is in equilibrium and coexists surprisingly with a Meissner state (complete shielding of an "external" magnetic field). The present observation not only reveals the evolution process of the spontaneous vortices in superconductor/ferromagnet hybrids, but it also sheds light on the role of the fractal disorder and structural heterogeneity on the vortex nucleation under the influence**

of Josephson superconducting currents.

There has been a long-standing interest in the interaction between superconductivity and ferromagnetism, both in terms of spintronics applications[12] and fundamental physics[2,3,11]. Since the internal exchange energy in ferromagnets is, in general, significantly higher than the condensation energy of superconductivity, a level of magnetic impurity of a few % can lead to a complete loss of superconducting order[14,15]. It has, however, been demonstrated that the coexistence of superconductivity and ferromagnetism can be achieved in some of the superconductor/ferromagnet (SC/FM) hybrid systems[3], in which the superconducting and ferromagnetic parts are electromagnetically coupled but are electrically separated by, for example, an intervening insulating layer. Recent developments in material fabrication and imaging techniques have enabled investigation of such purely magnetically coupled systems of both natural[16,17] and artificial[18–20] origin. As a result, various intriguing phenomena, such as domain wall superconductivity[19,20], domain Meissner state[17], and hysteretic vortex pinning[18], are confirmed to exist, which have been preceded or followed by theoretical studies[21–23].

Irrespective of these recent advances, many issues still remain to be solved concerning the interplay between superconducting and ferromagnetic orders. One of such issues is nucleation and expulsion processes of spontaneous vortices in SC/FM hybrids[4,6]. Spontaneous vortex phase is an exotic quantum state in which quantized vortices are formed in the absence of external magnetic field[24]. In the framework of the electromagnetic mechanism, a spontaneous vortex state is more favorable than the domain-like structure if the magnetization is sufficiently large[2]. Although the spontaneous formation of vortex-antivortex pairs has been recently observed at the surfaces of a ferromagnetic superconductor, $EuFe_2(As_{0.79}P_{0.21})_2$ (ref. 17), and SC/FM bilayers[25,26] by local scanning probe techniques, a definite observation of the spontaneous vortex state by bulk magnetization measurements has not been made[4,8,9,10].

Here, we tackle the above problem by utilizing superconducting fractal nanocomposites with magnetic inclusions. Over the last decade, a possible role of the quality of fractal (scale free) organization and related structural inhomogeneity in the development of superconductivity has been increasingly recognized[27–30]. Recently, we[31] reported superconducting properties of the $Mg/MgO/MgB_2$ nanocomposite prepared through the solid phase reaction between Mg and $B_2O_3$. In the nanocomposite, $MgB_2$ nanograins are distributed in a fractal manner with a fractal dimension of ~1.8 in the normal matrix. We[31] demonstrated that irrespective of the low volume fraction (~16 vol. %) of $MgB_2$ nanograins, the nanocomposite behaves as a bulk-like superconductor, i.e., zero resistivity, perfect diamagnetism, and strong vortex pinning. Hence, in the nanocomposite, a Josephson phase coherence is achieved throughout the volume of the sample due to the long-range proximity coupling among fractally distributed superconducting nanograins via quantum interference of Andreev quasiparticle[32]. In this work, we incorporated a small amount of Mn into the $Mg/MgO/MgB_2$ nanocomposites (see Methods for fabrication details). In these nanocomposites, α′-MnB, which is a soft to semi-hard ferromagnetic nanomaterial with a high Curie temperature $T_{FM}$ = ~540 K[11], has been found to exist as single domain magnetic particles in the normal matrix (Extended Data Figs. 1 and 2). The resistivity and magnetization measurements revealed that the samples containing 0, 0.24, 0.86 and 2.4 wt % Mn show the onset of superconductivity $T_c^{onset}$ at 38.5, 36, 34, and 23 K, respectively (Extended Data Fig. 3a), and a diamagnetic Meissner response at 2 K (Extended Data Fig. 3b). Hence, in these nanocomposites, the spin-flipping scattering will not be so strong as to induce a severe pair-breaking effect. This implies that the α′-MnB grains are physically and electronically separated from MgB$_2$ to suppress the exchange interaction. In what follows, we will mainly show the experimental results of the sample with 0.24 wt% Mn, unless otherwise indicated.

The 0.24 wt% Mn sample consists of MgO (~62 wt%), Mg (~28 wt%) and MgB$_2$ (~10 wt%), as demonstrated in a Rietveld analysis of the x-ray diffraction pattern shown in Fig. 1a. In this sample, boron, and hence MgB$_2$, was found to be distributed fractally with a fractal dimension of ~1.8 (Fig. 1b and Extended Data Fig. 4). We confirmed from the magnetization (*M*)-field (*H*) measurements that the sample shows a typical hysteresis loop of a ferromagnet at temperatures at 50 K (Fig. 1c) and that of a superconductor with strong flux pinning at temperatures below 10 K (Fig. 1c and Extended Data Fig. 5a). The lower ($H_{c1J}$) and upper ($H_{c2J}$) critical fields of this Josephson-coupled system are estimated to be ~70 Oe ((Extended Data Fig. 5b) and 54 kOe (Fig. 1d), respectively.

To investigate the formation process of possible spontaneous vortices in the present Josephson coupled system with magnetic inclusions, we carried out magnetization measurements under nominally zero applied field in the temperature *T* range from 300 K to 2 K (Fig. 2a). Before starting the measurements, we set the sample into three different remanent states at 300 K, i.e., the positive remanent (PR) state, negative remanent (NR) state, and demagnetized (DM) state (see Methods for details). The resulting remanent magnetizations for the PR, NR and DM states are +0.094, −0.094 and ~$1 \times 10^{-4}$ emu/cm³ (+0.16, −0.16 and ~$2 \times 10^{-4}$ $\mu_B$/Mn), respectively. As expected from the general temperature dependence of spontaneous magnetization in ferromagnets, the *M-T* curve of the sample in the PR (NR) state shows a gradual increase (decrease) with decreasing temperature from 300 K to ~40 K, whereas the *M* value of the DM state remains to be almost zero irrespective of temperature. At a temperature of ~32 K, however, the *M* value of the PR (NR) sample shows a steeper increase (decrease), then followed by a constant decrease (increase) in the temperature range below ~30 K. Considering that the zero-field resistivity of this sample becomes zero at temperatures below ~32 K (Fig. 1d), we suggest that the steeper increase (decrease) in *M* observed for the PR (NR) sample at temperatures below ~32 K results from the diamagnetism of the superconducting part of the sample in response to the stray field of the ferromagnetic grains, as has been often argued to occur in SC/FM hybrid systems[33–36]. However, the subsequent decrease in *M* with further decreasing temperature has hardly been reported in previous SC/FM systems.

Notably, during prolonged holding in zero field for additional hours at 2 K, the value of *M* for the PR (NR) state shows a continuous decrease (increase) (Fig. 2a) and eventually reaches a negative (positive) $M$ value (Fig. 2b). The time decay of $M$ can be well fitted to the stretched exponential function[37], $M(t) = A\exp(-(\frac{t}{\tau})^\beta) + C$, where $\tau$ is an effective decay time, $\beta$ is a stretching exponent representing the distribution of $\tau$, and $A$ and $C$ are fitting constants. The fitted values of $\tau$ and $\beta$ are 16–20 h and ~0.6, respectively, showing an extremely slow decay time constant with a broad distribution. The resulting negative (positive) magnetization for the PR (NR) state allows us to assume that the observed relaxation does not result simply from vortex creep[38] but from the confinement of vortices due to the Meissner currents, as will be discussed again later.

We also found that, as shown in Fig. 2c, this slowly-decaying negative (positive) value of $M$ for the PR (NR) state suddenly switches to a large positive (negative) value once a small external magnetic field is applied; for example, the value of $M$ for the PR state changes from −0.2 to 0.5 emu/cm$^3$ after applying an external field of $H$ = 0.1 Oe. Then, $M$ for the PR (NR) state starts to show a slight decrease (increase) during holding the sample under $H$ = 0.1 Oe, similar to the case of time decay observed in zero field. When the applied field of 0.1 Oe is switched off to zero, $M$ shows an abrupt jump again, followed also by slow time decay. By further repeating the on-and-off cycle of the applied field, the decay rate becomes slower and slower, converging to a constant $M$ value.

Qualitatively similar results were observed for the sample with 0.86 wt % Mn (Extended Data Fig. 6). As for the 0.24 wt % Mn sample, we further confirmed that field-induced magnetization jump and decay behaviors were observed as well when a small field was applied immediately after the system temperature reaches 2 K in zero-field-cooled (ZFC) mode (Fig. 2d). These results suggest that the application of a small external magnetic field can drive the system into a unique (quasi-)equilibrium state irrespective of the previous thermal history of the sample.

To confirm whether or not the system is in the (quasi-)equilibrium state after field application, we performed the isothermal magnetization measurements in the near-zero field region ($-20$ Oe $\leq H \leq 0$ Oe) (Fig. 3a, and see also Extended Data Fig. 7 for the 0.86 wt % Mn sample) One sees that the *M-H* data yield a good linear and reversible *M-H* relationship, confirming that these samples are virtually in equilibrium. Each *M-H* line not only has almost the same slope of $\frac{dM}{dH} = \sim -0.083 \pm 3$ emu/(cm$^3$Oe), which is close to that of a perfect diamagnetic state, i.e., $\frac{dM}{dH} = -\frac{1}{4\pi} = -0.080$ (cgs units), but it also has a unique vertical intercept depending on the initial remanent state. Thus, the magnetization in the equilibrium state can be represented by

$$M(H) = \chi_0 H + M_s \qquad (1)$$

where $\chi_0 \approx -\frac{1}{4\pi}$, and $M_s$ is the external-field-independent magnetization, i.e., the spontaneous magnetization. We should note that the $M_s$ values for PR and NR states ($|M_s|$ = 0.5 emu/cm$^3$ or 0.84 $\mu_B$/Mn) are even higher than the saturation magnetization at 50 K ($|M_s|$ = 0.44 emu/cm$^3$ or 0.74 $\mu_B$/Mn, see Fig. 1c). This large enhancement of $M$ most likely results from the nucleation and growth of the spontaneous vortices. When exceeding $H_{c1J}$, the external field begins to penetrate the material in the form of an "external" vortex. Then, we observed a nonlinear and irreversible $M$-$H$ relationship at fields larger than $H_{c1J}$, as shown in the inset of Fig. 3a.

Doria et al.[7,39] theoretically investigated the formation of vortex patterns induced by the magnetic inclusions embedded in a superconducting sphere in the absence of exchange coupling under the framework of the Ginzburg-Landau theory. These authors[7,39] demonstrated that a point like magnetic dipole deep inside a SC is under pressure by the Meissner current. Consequently, vortices can nucleate in triplets around the magnetic core, forming confined vortex loops (CVLs). As the dipole moment of the magnetic domain increases and draws near to the surface, the confinement of vortices becomes energetically less favorable. This thermodynamic modification allows a CVL to split into two vortex lines, leading to an alternative vortex state called an external vortex pair (EVP). Doria *et al.*[7] also showed that an external perturbation, e.g., an external current applied along the magnetic dipole, can induce the transition from CVL to EVP, accompanied by a jump in the magnetization from a negative to a positive value.

The above theoretical prediction satisfactorily fits the set of experimental results obtained in this work. As for the PR state, the *M* value continues to decrease when the sample is held under zero applied field at 2 K, eventually leading to negative *M* values (Fig. 2b). This decrease in *M* will represent the development of the Meissner state due to the formation of CVLs and the subsequent further confinement of vortex loops, as schematically shown in the insets of Fig. 2a,b. Here, we should recall that the present SC/FM nanocomposite is a disordered Josephson coupled system. This implies that the confinement process of the vortex loops will require substantial vortex rearrangements, which are impeded by activation energy barriers created in the nanocomposite. This will result in very slow and broadly distributed decay kinetics in *M*, as demonstrated in the relaxation data in Fig. 2b. Also, Doria *et al.*[7] have demonstrated that a CVL becomes unstable near to the surface of the superconductor and can be broken apart to form an EVP by external perturbation. Hence, the large jump in *M* observed at the moment of "on" and "off" of the applied field most likely results from the transition from a CVL to an EVP, as depicted in the insets in Fig. 2c, which is triggered by the induced eddy current created on the surface of the sample. It is also probable that the formation of EVP at the surface is followed by the growth of EVPs through the whole sample via the interconnection between neighboring magnetic inclusions[3] (Fig. 3c). After the development of the EVPs throughout the sample, however, some of the vortex loops will still remain unbroken and are to be confined to the magnetic grains due to the diamagnetic effect of Meissner currents. We suggest that a slow time decay in *M* observed during the field on/off periods results from the confinement process of these residual loops.

In general, the spontaneous vortex phase should be formed in the region where the internal field generated by the ferromagnetic moment is larger than $H_{c1}$ (ref. 5,6). From this follows that if the magnetic field is homogeneously distributed, a spontaneous vortex phase cannot, in principle, coexist with a Meissner state; that is, $H_{c1}$ would be zero, as indeed demonstrated, for example, in a ferromagnetic superconductor UCoGe (ref. 40). However, a spatially dispersed distribution of magnetic inclusions, as in the case of our sample, will not induce a homogeneous magnetic moment throughout the sample. Some regions may have an internal magnetic field higher than $H_{c1J}$, but others may not. This magnetic inhomogeneity will result in the occurrence of the anomalous equilibrium state in which a spontaneous vortex state coexist with a Meissner state. Thus, the present observations provide the evidence that a Josephson coupled SC/FM system with multiscale disorder induces intriguing interplay between superconducting and ferromagnetic orders that cannot be realized in the SC/FM hybrids with a homogeneous magnetic field distribution.

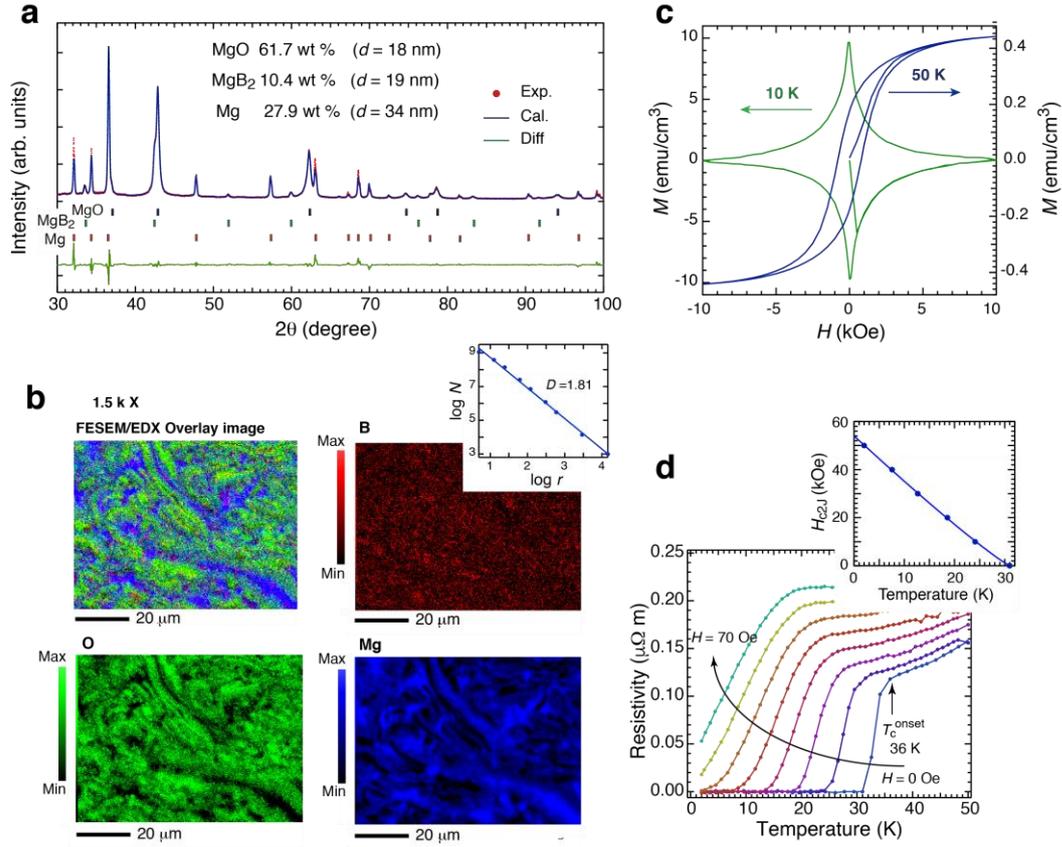

**Fig. 1 | Structural, electrical and magnetic properties of the Mg/MgO/MgB$_2$ nanocomposite with 0.24 wt% Mn. a,** Output from a Rietveld analysis of the X-ray diffraction pattern of the as-prepared powder sample. The weighted-profile $R$ value ($R_{wp}$) and the goodness-of-fit (GoF) are 8.4 % and 2.7, respectively. The red dots represent the measured intensity, the blue curve is the calculated profile, and the green curve is the difference between the measured and calculated intensities. The average crystalline sizes $d$ estimated from Scherrer equation are also shown. **b,** Field emission scanning microscopy/energy dispersive x-ray analysis (FESEM/EDX) of the surface of the sintered sample prepared by a spark plasma sintering method (magnification: 1.5k ×); Red = B, Green = O, Blue =Mg. The concentration of Mn was too low to be recognized visually. The inset shows the result of the box-counting analysis for boron distribution in the corresponding FESEM/EDX image. **c,** Magnetization hysteresis curves $M(H)$ measured at 10 and 50 K. **d,** Resistivity in different applied fields (from bottom right to top left) from 0 to 70 kOe in steps of 10 kOe. The inset shows the temperature dependence of the upper critical field of the Josephson coupled network $H_{c2J}$ determined from the transition temperature at zero resistivity for a given value of $H$. The solid line represents the fit of the function $H_{c2J}(T) = H_{c2J}(0)(1 - (T/T_c))^{1+\alpha}$. The fitted values of $H_{c2J}(0)$ and α are 54 kOe and 0.1, respectively.

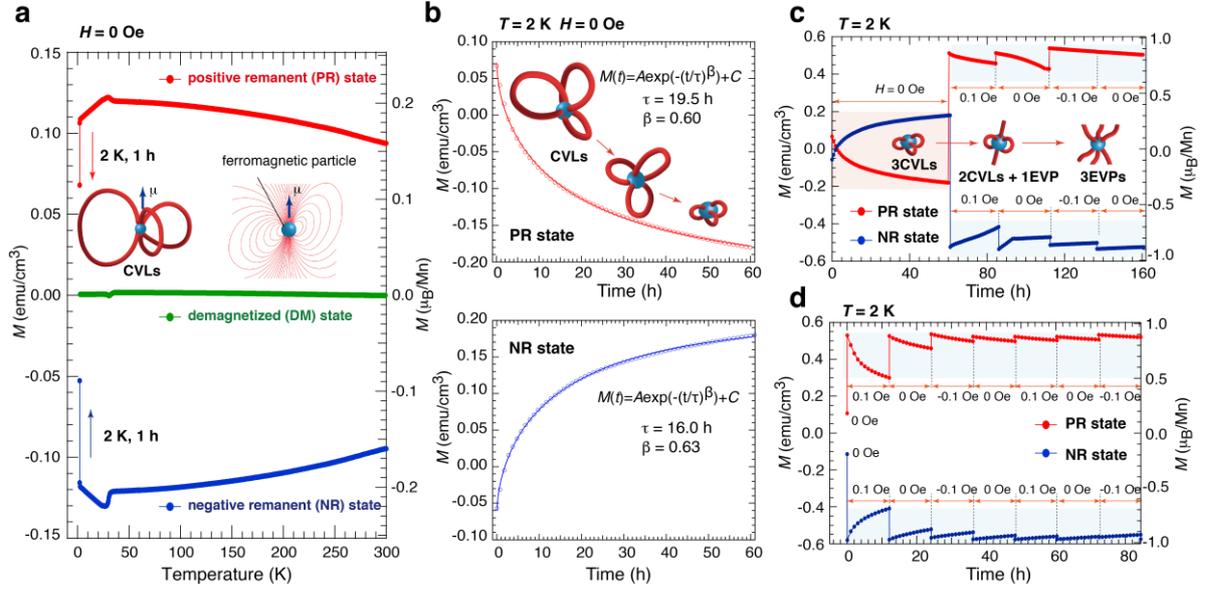

**Fig. 2 | Temperature and time-dependent magnetization of the Mg/MgO/MgB$_2$ nanocomposite with 0.24 wt % Mn in different remanent states. a,** Temperature dependence of *M* for the sample in the PR, NR and DM states taken under zero field during the cooling procedure from 300 to 2 K. At a temperature of 2 K, the sample is held for one hour under zero field, resulting in changes in *M* for the PR and NR states but not for the DM state. The insets illustrate the magnetic flux lines of a ferromagnetic particle in the normal state (right) and the resulting three confined vortex loops[7] (CVLs) nucleated in the superconducting state (left). **b,** Time-dependent decay of *M* under zero applied field at a temperature of 2 K for the PR (upper panel) and NR (lower panel) states after the zero-field-cooling procedure shown in **a**. The experimental data (open circles) are fitted to the stretched exponential function (solid line) $M(t)=A\exp(-(t/\tau)^\beta)+C$. The fitted values of $\tau$ and $\beta$ are also shown. The insets schematically illustrate the development of the vortex confinement of CVLs due to Meissner currents. **c,d,** Time-dependent change in *M* during the designated on/off cycles of the external magnetic field. In **c**, the external field of $H = 0.1$ Oe was applied after holding the sample in zero field at 2 K for 60 h, as also shown in **b**, whereas in **d**, the external field was applied immediately after the system temperature reached 2 K under zero applied field. The insets in **c** schematically depict the transition from three CVLs to three external vortex pairs[7] (EVPs) via two CVLs and one EVP during the field on/off cycles.

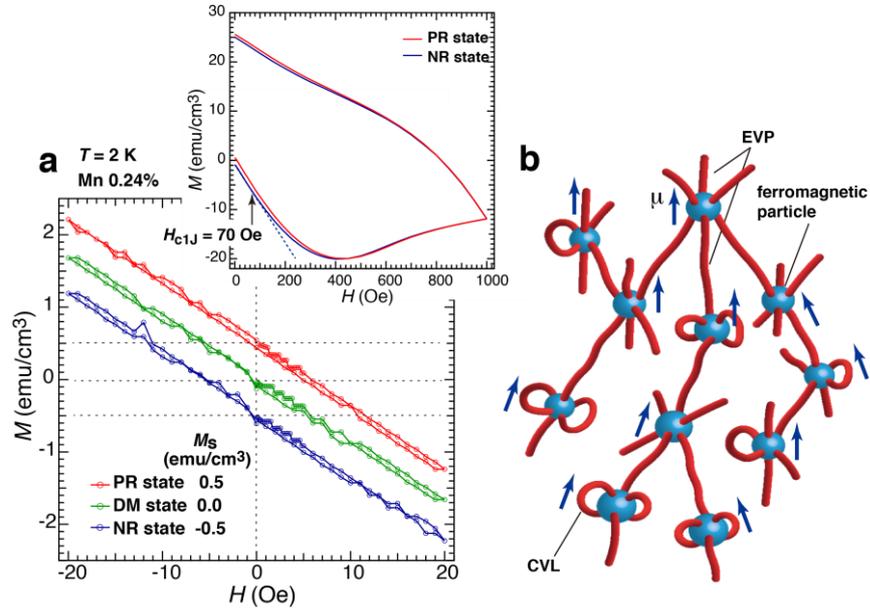

**Fig. 3| Low field *M(H)* data of the Mg/MgO/MgB₂ nanocomposites measured at a temperature of 2 K. a,** The linear and reversible *M(H)* relationship measured during the following field sweeping process: from 0.1 to 20 Oe, from 20 to -20 Oe, and from -20 to 0 Oe. The measurements were carried out after zero-field cooling from the respective remanent states. Before the measurements, the sample was held at 2 K under zero field for a certain period of time (normally several tens of minutes). We confirmed that the resulting *M(H)* relationship was not influenced by the previous thermal history of the sample. These lines have a common slope of ~−0.083±3 emu/cm³Oe. The values of the vertical intercept, or spontaneous magnetization $M_s$, for each *M(H)* line are given in emu/cm³. The insets show the low field part of the initial magnetization curve of the PR and NR states for *H* swept from 0.1 Oe to 1 kOe and then back to zero. **b,** Schematic illustration of a quasi-equilibrium state with a large $M_s$, value, where the neighboring magnetic particles are interconnected by EVPs throughout the sample and some residual CVLs are supposed to be present. The blue arrows indicate the direction of the magnetic dipole moment of the respective magnetic particles.

# Methods

**Sample preparation.** Pure Mg (99.9%), B2O3 (99.9%) and MnCO$_3$ (99.9%) powders were used as starting materials. The Mg/B$_2$O$_3$ mixture of molar ration of 5:1 was thoroughly mixed. We then added the MnCO$_3$ powder into the mixture powder of Mg and B$_2$O$_3$. The amount of Mn was varied by changing the weight ratio between MnCO$_3$ and Mg/B$_2$O$_3$ mixture from 0.2/100 to 25/100. The thus prepared mixture powder with a total weight of 2 g was put in a cylindrical alumina crucible. This crucible was located inside a larger rectangular alumina crucible, which was closed with a thick aluminum lid. This set of crucibles was placed in an electric furnace. The furnace was evacuated down to ~30 Pa, and was subsequently purged with argon. The temperature of the furnace was raised to 700 ˚C at a rate of ~10 ˚C/min and kept constant at 700 ˚C for 3 h under flowing Ar environment. During the heating procedure the following oxidation-reduction reactions are expected to occur:

$$MnCO_3 \rightarrow MnO + CO_2 \quad (>370\ ˚C)^{41},$$

$$3Mg + B_2O_3 \rightarrow 2B + 3MgO \quad (>500\ ˚C)^{42,43},$$

$$Mg + MnO \rightarrow Mn + MgO \quad (>~500\ ˚C),$$

$$Mg + 2B \rightarrow MgB_2 \quad (~700\ ˚C)^{43,44},$$

$$\text{Mn} + \text{B} \rightarrow \text{MgB} \ (\sim 700\ °\text{C})^{45}.$$

After the heating process, the furnace was naturally cooled to room temperature, yielding black powders in the inner crucible. The Mn concentration levels in these samples were determined by induction coupled plasma-atomic emission spectroscopy. Then, the powders were sintered into 15 mm diameter disk-shaped pellets using a spark plasma sintering (SPS) method for electrical and magnetic measurements. The collected black powders weighing about 2.0 g were loaded into a 15 mm diameter graphite die and were processed using a SPS system (SPS-725K, Fuji Electronic Ind. Co., LTD., Fujimi, Japan). In SPS, sintering is realized by subjecting the green compact to arc discharge generated by a pulsed electric current. An electric discharge process takes place on a microscopic level and accelerates the sintering processes accompanied by material diffusion. One of the most pronounced features of this technique is that the small grain size can be maintained while achieving full densification, enhancing the connectivity between grains through the minimization of the undesirable grain growth and the creation of clean grain boundaries[46]. Thus, the post SPS treatment is a key process for preparing a densified nanocomposite with clean interfaces and high connectivity. The pulsed electric current (5000 A) was passed through the sample under dynamic vacuum (~50 Pa) while a 100

MPa uniaxial pressure was applied. The heating rate was 60 ˚C/min up to 650 ˚C. During the experiment, the temperature, applied pressure, displacement (shrinkage), and environmental pressure were recorded. The onset temperature of the densification process was found around 300 ˚C. Accordingly, a cylindrical bulk sample with a diameter of 15 mm and a length of ~4 mm was obtained. The bulk density of typically 2.435 g/cm$^3$ was obtained by dividing the weight in air by the geometric volume of the specimen. The thus obtained SPS-treated samples were cut into suitable shapes for later characterization.

**X-ray diffraction measurements**

X-ray diffraction (XRD) patterns of the as-prepared powder samples were obtained with a diffractometer (SmartLab, Rigaku Corporation, Osaka, Japan) equipped with a sealed tube X-ray generator (a copper target; operated at 40 kV and 30 mA). Rietveld refinement of XRD patterns was performed to accurately determine the phases and their quantitative compositions using Rigaku's PDXL software with the Whole Pattern Powder Fitting[47] (WPPF) method, connected to the PDF2 database of International Centre for Diffraction Data (ICDD, Newtown Square, USA). The refinements were carried out by checking the fitting quality of the powder patterns by means of two parameters, i.e., the goodness-of-fit indicator GoF and the weighted *R*-factor of Rietveld refinement $R_{wp}$.

**FESEM observation**

Field-emission scanning electron microscopy (FESEM) and energy dispersive X-ray (EDX) spectroscopy were conducted on cross-sections of the samples cut from the SPS-treated samples using a scanning electron microscope (JSM-7100F, JEOL, Tokyo, Japan) with an EDX spectrometer. Before the FESEM measurements, the surface of the sample was polished with a emery paper (4000 grit).

**Fractal analysis**

To evaluate the fractal dimension for the distribution of $MgB_2$ nanograins in the composite, box counting was applied on FESEM/EDX mapping images of boron K$\alpha$. Firstly the original EDX mapping images were converted to an 8-bit format and then finally transformed to a binary format. Image processing and fractal analysis were carried out in the ImageJ 1.52a software (NIH, Bethesda, USA). Box counting technique is consisted of covering the slice image with voxels (boxes in 2D). The fractal dimension or box-counting dimension *D* is obtained by the following relationship[48]:

$$D = \frac{\log(N)}{-\log(r)}$$

where $r$ is variable box dimension in pixels and $N$ is the minimum number of boxes needed to encompass the whole object containing the boron K$_\alpha$ signals. If any linear region is observed in the curve $\log(N)$ versus $\log(r)$, the fractal dimension is equal to the slope with a negative sign of the curve.

**Electrical resistivity and magnetoresistivity measurements**

The temperature and magnetic field dependence of the electrical resistivity was monitored using a commercial superconducting quantum interference device (SQUID) magnetometer (MPMS-XL, Quantum Design, San Diego, USA) in the temperature interval 2–300 K and in magnetic fields up to 7 T. For transport measurements, we employed a dc four-probe technique using a square cuboid sample with a size of $2 \times 2 \times 10$ mm$^3$ where four gold wires are attached to the sample using silver paste. In this work, the longitudinal magnetoresistivity was measured; that is, the magnetic field was aligned along the direction of current flow.

**Magnetic susceptibility measurements**

The DC magnetization measurements were carried out for a square cuboid shape sample with dimensions of $1 \times 1 \times 5$ mm$^3$ using a commercial SQUID magnetometer (MPMS-XL, Quantum Design, San Diego, USA) equipped with the reciprocating sample option (RSO). The positive (negative) remanent state was obtained by applying the positive field of $H$ = 10 (–10) kOe for 5 min at 300 K. The resulting positive (negative) remanent magnetization at zero field was $M$ = ~+0.94 (~–0.94) emu/cm$^3$ for the sample with 0.24 wt % Mn. To obtain the demagnetized state, the sample was first exposed to $H$ = 10 kOe, and then the field was successively reduced by ~15 % whilst alternating its polarity. The residual remanent magnetization after the above demagnetization procedure was typically $M = \sim \pm 1 \times 10^{-4}$ emu/cm$^3$. The temperature-dependent magnetization in the positive (negative) remanent state in zero field was taken with a cooling rate of 1 K/min in the temperature range from 300 to 2 K

The external magnetic field, if needed, was applied along the long side of the cuboid sample mentioned above. In this configuration, the resulting effective demagnetization factor is estimated to be 0.05 (ref. 49), and the magnetizations and susceptibilities were not corrected for this small demagnetization factor. To estimate the volume magnetization *M* we used the bulk density (2.435 g/cm$^3$ for the sample with 0.24 wt % Mn).

In addition to the conventional *M(H)* hysteresis loops, we obtained first-order reversal curves[50,51] (FORCs) to gain qualitative information about the presence of magnetic materials with varying domains states and the presence or absence of magnetostatic interactions[50–53]. The FORC method is particularly sensitive to irreversible switching processes during magnetization reversal, giving rise to distributions of magnetic characteristics, such as the coercivity distribution and the amount of interparticle magnetic interaction. The measurement of a FORC begins with the saturation of the sample by a large positive applied field, e.g., *H* = 10 kOe. The field is then ramped down to a reversal field $H_R$. The FORC consists of a measurement of the magnetization as the applied field is increased from $H_R$ back to saturation, tracing out a FORC. A family of FORC's is measured at different $H_R$, with equal field spacing, thus filling the interior of the conventional major hysteresis loop. The FORC distribution ρ is then defined by a mixed second-order-derivative[50,51],

$$\rho(H_R, H) \equiv -\frac{1}{2}\frac{\partial^2 M(H_R, H)}{\partial H_R \partial H},$$

which is well defined only for $H \geq H_R$. This eliminates the purely reversible component of the magnetization, indicating that any nonzero ρ corresponds to irreversible switching processes. Generally, FORC diagrams are presented in terms of the bias field $H_b$ and coercivity $H_c$ distributions using the transformations[50,51]: $H_b = \frac{H+H_R}{2}$ and $H_c = \frac{H-H_R}{2}$. Accordingly, the resulting contour plot maps out the coercivity and bias field distributions of the constituent magnetic particles.

## Data availability

The data that support the findings of this study are available from the corresponding author upon reasonable request.

**Acknowledgments**   This research was carried out under the joint research program of Molecular Photoscience Research Center, Kobe University with Proposal No. R0106. A part of this research was supported by Nanotechnology Platform Program (Molecular and Materials Synthesis) of the Ministry of Education, Culture, Sports, Science and Technology, Japan with Grant No. JPMXP09S19NR0023 and JPMXP09S19MS1063b. We are grateful to T. Ogawa and K. Ohishi for their support in the high-temperature magnetization measurements. T. U. acknowledges support from the Nippon Sheet Glass Foundation for Materials Science and Engineering. Spark plasma sintering was performed at Kojundo Chemical Laboratory Co., Ltd., Saitama, Japan.

**Author contributions** T. U. conceived and designed the experiments. N. T., T. U., Y. S., T. S., and H. O. performed the experiments. T. U. and N. T. analyzed the data and prepared the manuscript, with input from all authors.

**Competing interests**   The authors declare no competing interests.

**Additional information**

**Correspondence and requests for materials** should be addressed to T. U.

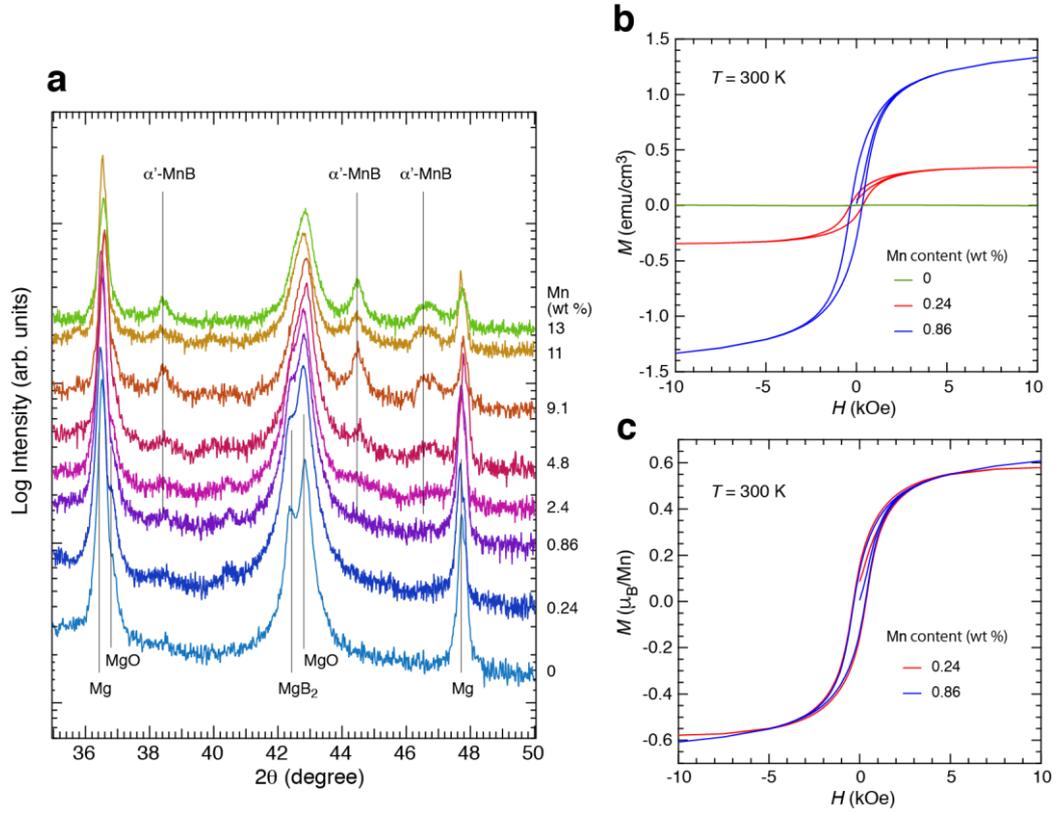

**Extended Data Fig. 1 | Structural and magnetic properties of the Mg/MgO/MgB$_2$ nanocomposites with different Mn contents measured at a temperature of 300 K. a,** X-ray diffraction patterns of the nanocomposites with Mn contents ranging from 0 to 13 wt %. The diffraction peaks attributed to α′-MnB (ref. 11) can be recognized for the samples with Mn content larger than 0.86 wt %. **b,** $M(H)$ loops for the samples with 0, 0.24, and 0.86 wt% Mn. Here, the unit of magnetization is emu/cm$^3$. **c,** The same data as in **b**, but presented in the unit of $\mu_B$/Mn for the samples with 0.24 and 0.86 wt % Mn. The two $M(H)$ loops are almost identical, confirming that the observed magnetization results from Mn.

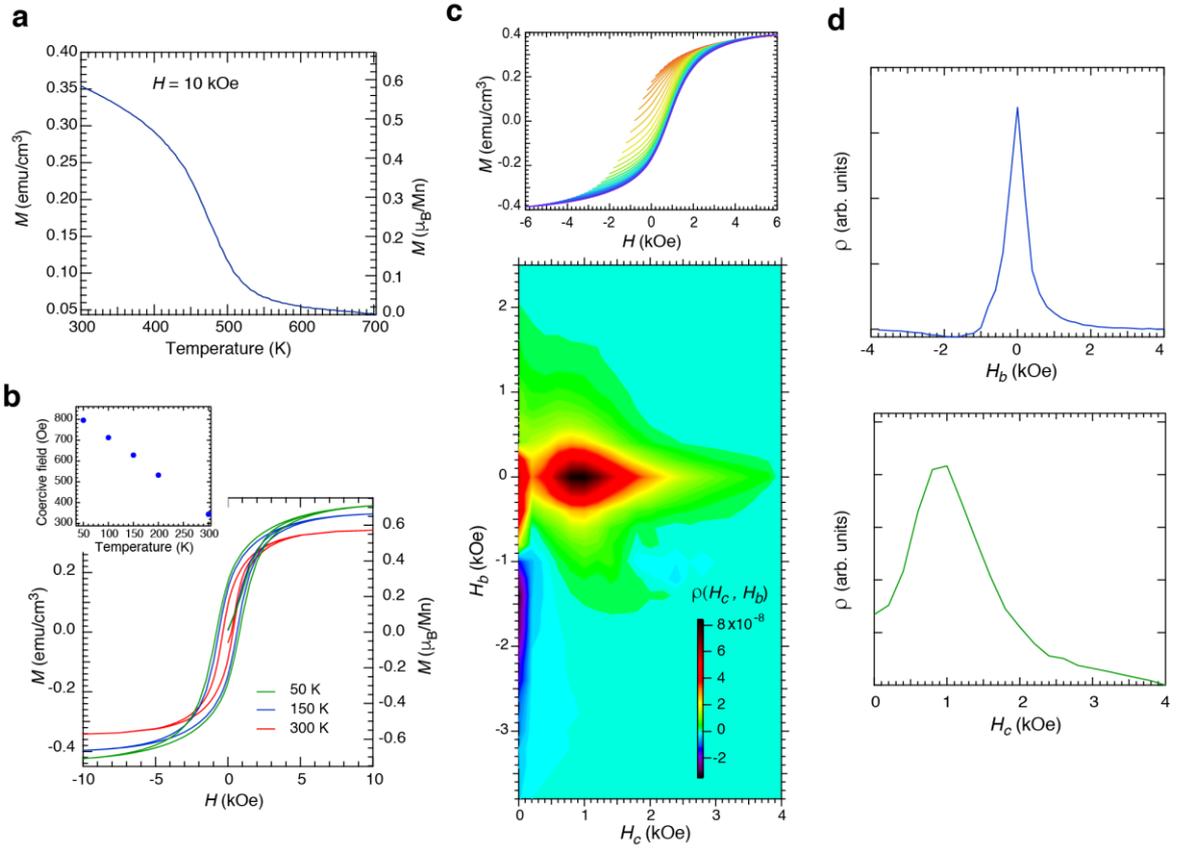

**Extended Data Fig. 2 | Ferromagnetic properties of the Mg/MgO/MgB$_2$ nanocomposites with 0.24 wt % Mn. a,** Temperature dependence of magnetization measured under applied field of 10 kOe. The ferromagnetic-to-paramagnetic transition occurs at $T_{FM}$ = ~540 K. **b,** A series of $M(H)$ loops measured in the temperature range from 50 to 300 K. The inset shows variation of the coercive field $H_c$ as a function of temperature $T$. The observed value of $T_{FM}$ along with the linear temperature dependence of $H_c$ is fully consistent with those reported for nanoscale ferromagnetic α′-MnB (ref. 11). **c,** A family of first-order reversal curves (FORCs, upper panel) and the corresponding FORC distribution plotted using the $H_c - H_b$ coordinate (lower panel). The FORC distribution ρ has a feature in the form of a peak stretching along $H_c$ and centered about $H_b = 0$. The large aspect ratio of the peak with a teardrop shape demonstrates that the embedded ferromagnetic particles are largely noninteracting[52–54]. One also notices that the negative region rises to a ridge along the negative $H_b$ axis. This is also a feature of noninteracting single domain nanoparticles[52,53]. **d,** Projections of the FORC distribution onto $H_b$ (upper panel) and $H_c$ (lower panel) presented in **c**, confirming small spread in $H_b$ versus large spread in $H_c$. This type of feature is typical of assemblies of single domain nanoparticles[52–54].

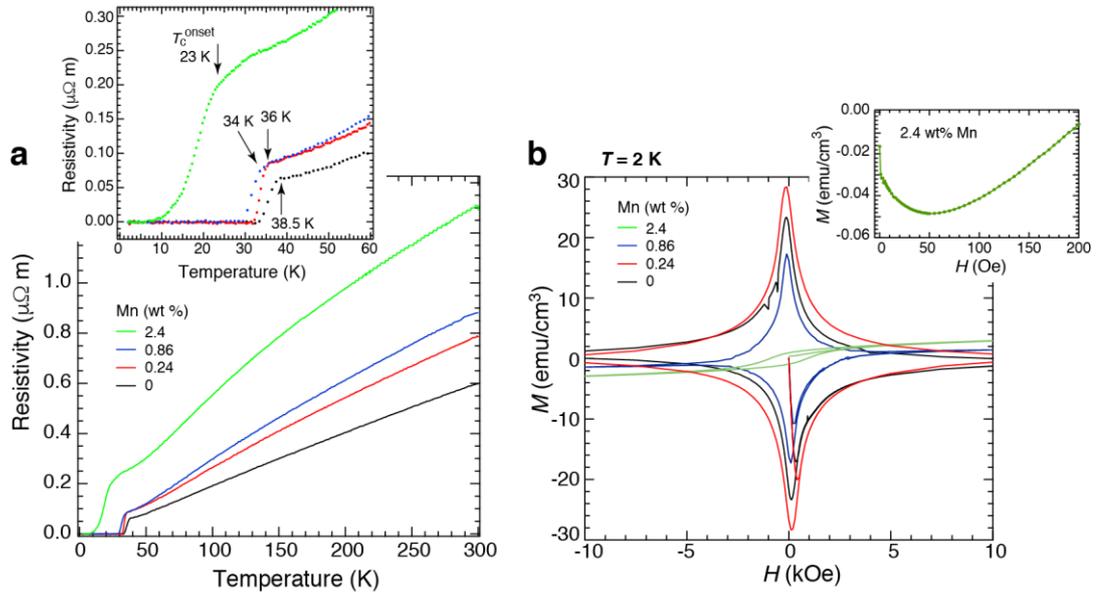

**Extended Data Fig. 3 | Resistive and magnetic properties of the Mg/MgO/MgB$_2$ nanocomposites with different Mn contents. a,** Temperature dependence of resistivity. The inset shows an enlarged plot around the superconducting transition region. **b,** $M(H)$ loops measured at 2 K. The inset shows the initial $M(H)$ curve of the sample with 2.4 wt % Mn in applied fields below 200 Oe. The samples with 0, 0.24 and 0.86 wt% Mn show a typical hysteresis loop of superconductors with strong pinning. The sample with 2.4 wt % Mn exhibits a ferromagnetic hysteresis loop, although a small diamagnetic response was observed in the initial $M(H)$ curve in the field region below 50 Oe, as shown in the inset.

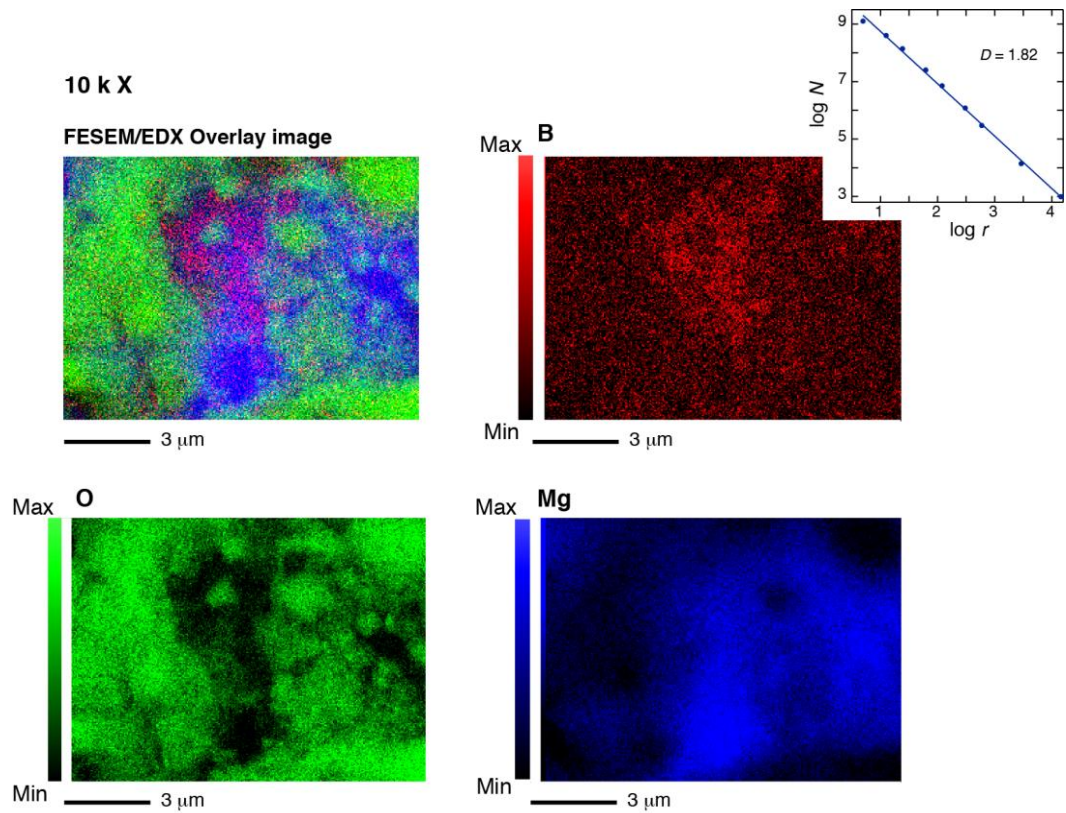

**Extended Data Fig. 4 | FESEM/EDX images at higher (10 k ×) magnification than those shown in Fig. 1b.** Red = B, Green= O, Blue = Mg. This image was obtained by enlarging one of the B-rich regions shown in Fig. 1b where the micrometer-sized red spots exist. One sees that the central micrometer-sized red region consists of several smaller sub-micrometer-sized red spots, yielding a fractal-like dendritic morphology. The inset shows the result of the box-counting analysis for boron distribution shown in this FESEM/EDX image. The box-counting dimension $D$ is calculated to be 1.82, which is almost identical to that obtained from the lower magnification image shown in Fig. 1b. This allows us to confirm the fractal-like distribution of $MgB_2$ nanograins. The distribution of Mn was not clearly recognized because of its low concentration.

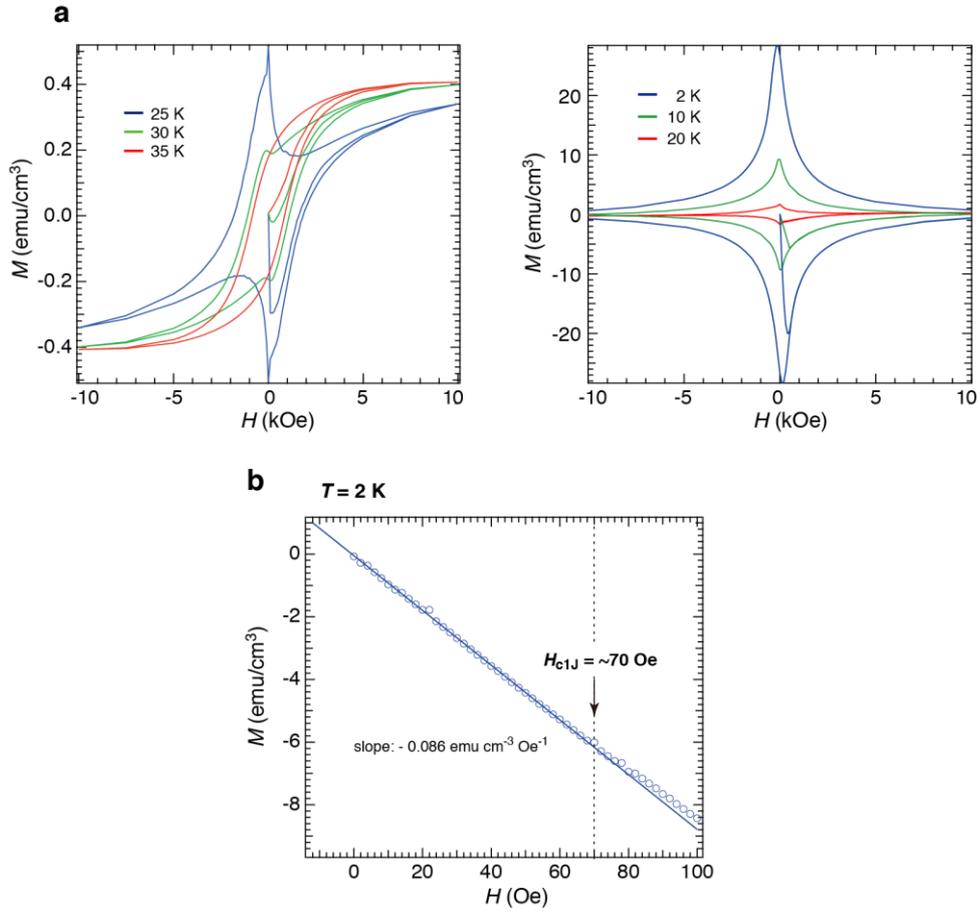

**Extended Data Fig. 5 | Zero-field cooled (ZFC) Magnetization-field *M(H)* responses of the demagnetized sample with 0.24 wt % Mn. a,** A series of *M(H)* hysteresis loops measured at different temperatures in the magnetic field range −10 kOe ≤ *H* ≤ 10 kOe. **b,** The initial *M(H)* curve in the magnetic field range from 0 to 100 Oe. The lower critical field of the Josephson-coupled superconducting network ($H_{c1J}$) can be defined as the deviation from the linear *M(H)* relationship, yielding the value of $H_{c1J}$ = ~70 Oe. The slope of the linear region is $-0.086$ emu/cm³Oe. This value is near to that expected from perfect diamagnetism, i.e., $-1/4\pi = -0.080$ (cgs units).

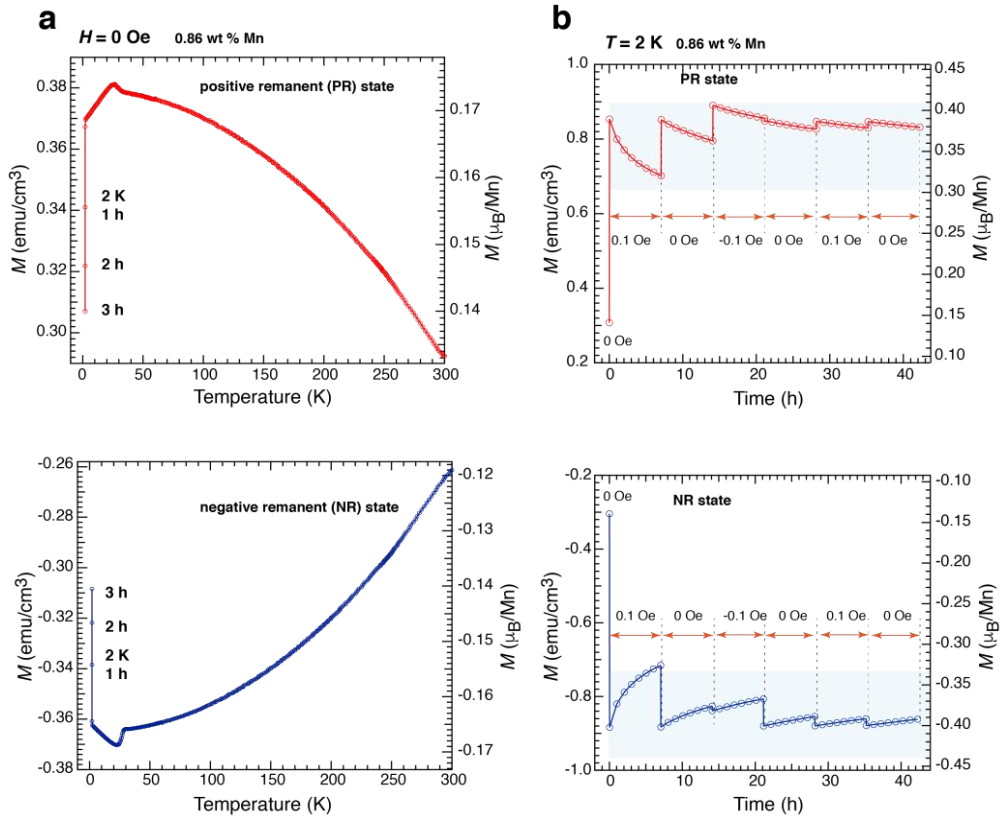

**Extended Data Fig. 6 | Temperature and time-dependent magnetization of the Mg/MgO/MgB$_2$ nanocomposite with 0.86 wt % Mn in different remanent states.** **a,** Temperature dependence of $M$ of the sample in the positive (upper panel) and negative (lower panel) remanent states taken under zero applied field during the cooling procedure from 300 to 2 K. Then, the sample was held at 2 K for additional 3 h under zero field, showing a substantial change in $M$ for a prolonged holding time. **b,** Time-dependent change in $M$ of the sample in the positive (upper panel) and negative (lower panel) remanent states during the designated on/off cycles of the external magnetic field. The measurements were carried out at a temperature of 2 K after zero-field-cooling procedure shown in **a**. Field-induced magnetization jump and decay behaviors were clearly observed. These experimental results are qualitatively similar to those obtained for the sample with 0.24 wt % Mn shown in Fig. 2.

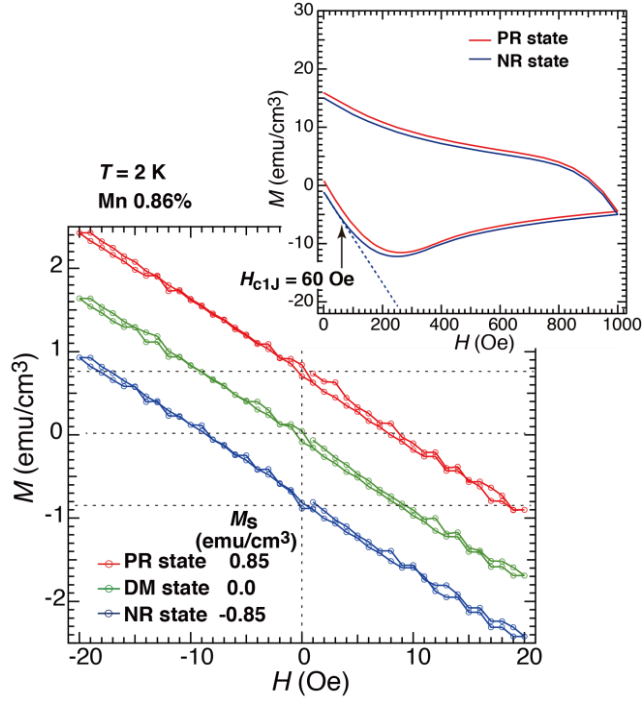

**Extended Data Fig. 7 | Low field *M*(*H*) data of the Mg/MgO/MgB$_2$ nanocomposite with 0.86 wt %Mn measured at a temperature of 2 K. a,** The linear and reversible *M*(*H*) relationship measured during the following field sweeping process: from 1 to 20 Oe, from 20 to -20 Oe, and from -20 to 0 Oe. The measurements were carried out after zero-field cooling from the respective remanent states. Before the measurements, the sample was held at 2 K under zero field for a certain period of time (normally several tens of minutes). We confirmed that the resulting *M*(*H*) relationship was not influenced by the previous thermal history of the sample. These lines have a common slope of ~–0.083±3 emu/cm$^3$Oe. The values of the vertical intercept, or spontaneous magnetization $M_s$, for each *M*(*H*) line are given in emu/cm$^3$. The insets show the low field part of the initial magnetization curve of the PR and NR states for *H* swept from 0.1 Oe to 1 kOe and then back to zero. These experimental results are qualitatively similar to those obtained for the sample with 0.24 wt % Mn shown in Fig. 3.